\documentclass{aa}  

\usepackage{graphicx}
\usepackage{subfig}
\usepackage{txfonts}

\usepackage{hyperref}
\usepackage{threeparttable}
\usepackage{caption}
\usepackage{stfloats}
\usepackage{placeins}
\usepackage{multicol}

\begin{document} 

\title{The spatial distribution of dwarf and giant galaxies in and around Virgo cluster}


   \author{Nelvy Choque-Challapa\inst{1,2}
       \and Rory Smith\inst{2,5}
        \and  Iv\'an Lacerna\inst{1}
        \and J. Alfonso L. Aguerri\inst{3}
        \and Daniela Palma\inst{1}
        }

   \institute{Instituto de Astronom\'ia y Ciencias Planetarias, Universidad de Atacama, Copayapu 485, Copiap\'o, Chile.\\
             \email{nelvy.choque@usm.cl}
            \and Departamento de F\'isica, Universidad T\'ecnica Federico Santa Mar\'ia, Av. Vicuña Mackenna 3939, 8940897, San Joaqu\'in, Santiago, Chile.\\
            \and Instituto de Astrof\'isica de Canarias, E-38200 La Laguna, Tenerife, Spain.\\
            \and Departamento de Astrof\'isica, Universidad de La Laguna, Avenida Astrof\'isico Francisco S\'anchez s/n, 38206 La Laguna, Tenerife, Spain.\\
            \and Millenium Nucleus for Galaxies (MINGAL).}

   \date{Received -, -; accepted -, -}

  \abstract
   {The Virgo cluster is one of the closest clusters to us where we can further study the evolution of galaxies, with several infalling substructures and several filaments around it have been reported. Therefore, it makes this cluster and its surrounding an interesting place to study the spatial distribution of the population of dwarf and bright giant galaxies. We analyse the dwarf fraction (DF) in different regions of the cluster, inside the virial radius, in its surrounding area, and in the filamentary structure surrounding it using available catalogues with the aim of measuring whether the DF changes in different environments.  Although the total dwarf fraction within the cluster is $\sim$ 0.8, significant local variations are measured throughout the cluster; there are regions with a relatively higher concentration of giant or dwarf galaxies.  The fact that Virgo is embedded in a rich environment surrounded by several filaments that feed the cluster with new substructures could imply changes in the DF locally. When we analyse the DF variation at further distances from the cluster we observe some regions with few or no giant galaxies at all, with a locally DF ranging from 0.8 - 1.0. Additionally, when comparing the dwarf fraction in different environments, overall, the DF is larger in regions further away from denser regions such as the Virgo cluster and its filamentary structure surrounding it. When comparing the filament and the cluster area, the dwarf fraction is slightly higher in the filaments, but from filament to filament, the DF changes depending on the presence of groups. }


   \maketitle
%

\section{Introduction}

Galaxies are distributed in a complex structure in the so-called cosmic web, from dense regions in nodes, which are connected through filaments, to voids in the lowest-density regions \citep{Haynes1986}. In the densest regions, galaxy clusters form and grow through the accretion of galaxies from the outskirts. Therefore, in these different environments galaxies are subject to different processes that shape their evolution.  An examination of the galaxy population in different environments facilitates our understanding into the processes that shape galaxy evolution, for example by studying the Luminosity Function (LF) in different environments \citep[e.g.,][]{Sabatini2003,Agulli2017}. For instance, the steepness faint-end slope of the LF gives us some indication of the abundance of less massive galaxies compared to the abundance of bright galaxies. Quite related and complementary to the LF is also the use of the dwarf-to-giant ratio in clusters \citep[DGR,][]{Ferguson1992,Sabatini2005,Popesso2005}.

The DGR actually gives us information about the distribution of dwarf galaxies in clusters, as a hint to explain their evolution, and gives some insight into the hierarchical assembly history of galaxies in the large-scale picture. Several studies have quantified the DGR in several clusters and have found, for example, that the DGR increases with a larger cluster-centric radius \citep[e.g.,][]{Barkhouse2009, Rude2020}. In any case, any change in the DGR with the cluster-centric distance could be due to the variation of giants, dwarfs, or both, as pointed out in \cite{SanchezJanssen2008}. For example, the decrease in the DGR when going to the innermost region of the cluster could be due to galaxy disruption, as gravitational tidal interactions become more frequent \citep{Trujillo2002,Aguerri2004,Rude2020} or to an increase in giants due to mass segregation. 
Similarly, the dwarf-to-giant ratio has also been studied as a function of some cluster properties, for example, to determine whether there is a correlation with the cluster mass, velocity dispersion, and X-ray luminosity \cite[e.g.,][]{Popesso2005}. The results on this aspect vary widely, with some works finding a correlation while others finding insignificant correlations with the cluster parameters \citep[e.g.,][]{Popesso2005, Bildfell2012,Hashimoto2018, ChoqueChallapa2021}. In addition to the connection of DGR to the environment, some observational studies have also quantified how it changes over time and found an evolving DGR, that is, a higher DGR at higher redshifts \cite[e.g.,][]{deLucia2007, Bildfell2012} supporting the hierarchical merging scenario where a higher fraction of smaller galaxies are expected at higher redshift.

Located at a distance of approximately 16.5 Mpc \citep{Mei2007}, and mass $\sim$ 7 $\pm$ 0.4 $\times$ $10^{14}$ M$\odot$,  \cite{Karachentsev2014}, the Virgo cluster is one of the closest massive galaxy clusters. Due to its proximity, this is a well-studied galaxy cluster, with several photometric and spectroscopic catalogues available to study galaxy evolution.  Within this system the mass assembly is still ongoing. Observations indicate the presence of numerous substructures that are currently falling into the cluster \citep{Binggeli1985,Binggeli1993,Bohringer1994,Lisker2018}. Furthermore, an extensive network of red filaments encircles the cluster \citep{Kim2016, Chungkim2021, Castinagni2022}. Consequently, this region, encompassing both the cluster and its surroundings, presents a compelling site for investigating its galaxy populations.

The properties of the population of bright giant and dwarf galaxies in Virgo have been widely studied \citep[e.g.,][]{Binggeli1985,Aguerri2005, Lisker2007, Jans2016, Kim2024}, and there is also an increasing number of studies on the filamentary structure around the cluster \citep{Chung2021, Castinagni2022, Zakharova2024}. It also becomes interesting to analyse their spatial distribution in and around this massive cluster.

Regarding the distribution of galaxies in the Virgo cluster, \cite{Sabatini2003} analysed a sample of dwarfs with magnitudes -14 $\leq$ M$_B$ $\leq$ -10 mag and giants with M$_B$ $\leq$ -19 mag and measured that the dwarf-to-giant ratio remains flat with distance from M87 central galaxy,  with a DGR having a median value of $\sim$  20. Similarly, \cite{Roberts2004} performed an analysis of the Millennium Galaxy strip, which includes the Virgo cluster, and found similar DGR value for the cluster, while the ratio along the strip ranges from 0.7 to, at most, 6 which corresponds to a flat luminosity function. One caveat when comparing this DGR value in the same cluster and also with other clusters is the different definitions of dwarfs and giant galaxies that studies use, and also some possible contamination from background galaxies when redshift is not available \citep{Hashimoto2018}. In addition to the DGR, \cite{Sabatini2005} analysing the dwarf galaxy distribution in the Virgo cluster found that a significant fraction of dwarf galaxies are not bound with giant galaxies. This apparently non-associated population of dwarfs corresponds to 40\%. Similar results were already pointed out in \cite{Ferguson1992}. 

While the majority of studies have been focused on the `global’ Dwarf-to-Giant Ratio  within the virial  radius (R$_{200}$\footnote{R$_{200}$ corresponds the radius within which the mean density is 200 times the critical density of the Universe.}), it is also of significant interest to analyse this ratio in different regions of the cluster and at larger distances, and see what we can learn from this. For example, the DGR could tell us more
about the assembly history of clusters by assessing how it changes as functions of cluster-centric and filament distance, and within substructures in and around a galaxy cluster. Moreover, one could also learn about the evolution of the dwarf galaxies in their falling into the clusters. Therefore, in
this analysis, we delve more into this aspect in the Virgo cluster and its surrounding environment analysing the spatial distribution of its dwarf and giant bright galaxies at different scales.

The paper is organised as follows. Section \ref{sec:data} describes the data sample we use in this analysis. The spatial distribution analysis of our target sample is presented in Sect. \ref{sec:results}. In Sect. \ref{sec:summary}, we discuss and summarise our main results.

Throughout this work, we adopt a Lambda Cold Dark Matter ($\Lambda$CDM) cosmology with $\Omega_m$ = 0.3, $\Omega_{\Lambda}$ = 0.7, and H$_0$ = 70 km s$^{-1}$ Mpc$^{-1}$, a distance module in Virgo, m - M = 31.09 \citep{Jerjen2003, Mei2007}, and adopted M87 galaxy as the centre of the Virgo cluster.


\section{Data}
\label{sec:data}

In this analysis, we make use of different available catalogues that cover different areas around the Virgo cluster providing us with photometric and spectroscopic information.  We use the Extended Virgo Cluster Catalogue \citep[EVCC;][]{Kim2014} which is based on the Sloan Digital Sky Survey (SDSS) Data Release 7, and is an extension of the Virgo Cluster Catalogue \citep[VCC;][]{Binggeli1985}. The EVCC catalogue covers an area of 725 deg$^2$ or 60 Mpc$^2$ reaching out $\sim$ 3.5 times the virial radius of Virgo (but mainly in the north hemisphere).  Additionally, this catalogue contains 1324 galaxies with radial velocities and SDSS $r$, $i$, $g$, and $u $ photometry, where the apparent magnitudes go down to m$_r$ $\sim$ 20 mag in the $r$-band.

Similarly, we make use of the \cite{Castinagni2022} galaxy catalogue, which covers a huge area in the north hemisphere surrounding the Virgo cluster (extending out $\sim$ 12 virial radius). This catalogue combines galaxies from Hyperleda \citep{Makarov2014}, NSA \citep[Nasa Sloan Atlas;][]{Blanton2011}, ALFALFA \citep{Haynes2018}, and NED-D \citep{Steer2017} containing a total of 6780 galaxies with radial velocity of 300 $<$ $v_r$ $<$ 3300 km s$^{-1}$. In order to have a homogeneous photometric sample for our analysis, we selected only galaxies coming from NSA and retrieved their corresponding $r$-band photometry.  Both the EVCC and the NSA-Castignani catalogues have a completeness magnitude, m$_r$,  up to $\sim$ 17 mag.

\begin{figure*}
    \centering
    \includegraphics[width=1\linewidth]{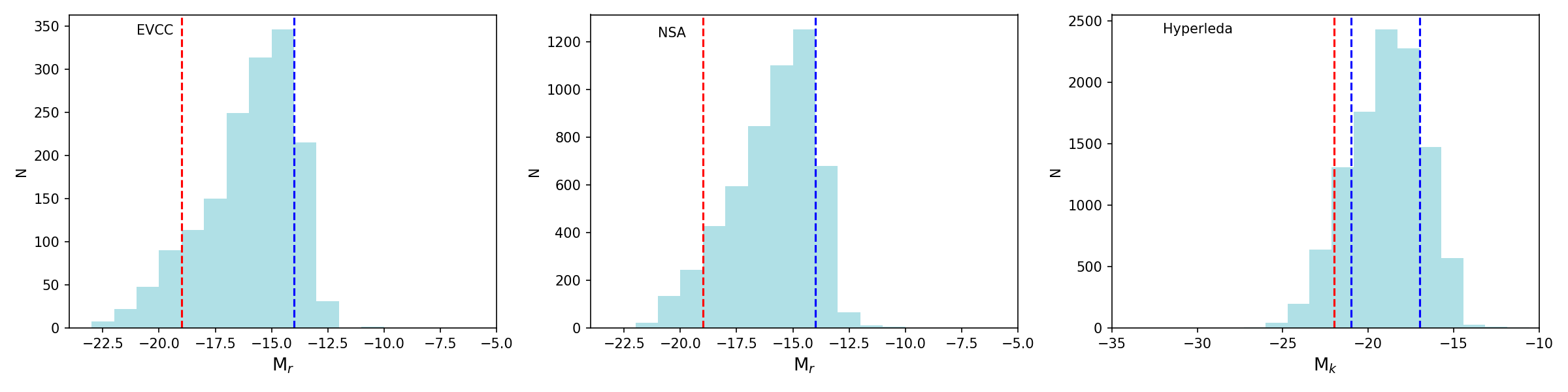}
    \caption{Absolute magnitude range of galaxies in the EVCC ($r$-band, left panel), NSA-Castignani ($r$-band, middle panel), and Hyperleda ($K$-band, right panel) catalogues. Red dashed  line indicates the threshold limit that we use to separate giant (left side) from  dwarf (right side) galaxies in the EVCC and NSA catalogues. Rightmost blue dashed line marks the limit imposed on dwarf galaxies to ensure the completeness of magnitudes of each catalogue. The blue dashed lines in the third panel indicates the M$_k$ range to select dwarf galaxies in the Hyperleda catalogue. Note that absolute magnitudes are measured at the distance of Virgo.}
    \label{fig:fig1}
\end{figure*}

Furthermore, given that both catalogues only cover the north hemisphere, we also make use of the catalogue described in \cite{Makarov2011}, which gathers data from the Hyperleda \citep{Paturel2003} and the NASA Extragalactic Database (NED) for photometry and radial velocities.  This catalogue covers not only the north area around Virgo but also a large portion in the south hemisphere (around 40 degrees) and contains $\sim$ 10000 galaxies with radial velocities  $<$ 3500 km s$^{-1}$ (z $<$ 0.01). It provides $K$-band photometry down to a magnitude m$_K$  $\sim$ 20 mag (but with a completeness magnitude m$_k$ up to $\sim$ 13 mag). In Fig. \ref{fig:fig1} we show the coverage of each catalogue and the limit of the magnitude range we are using in this analysis.

\begin{table*}
	\centering
	\caption{Number of dwarf and giant galaxies in the EVCC catalogue (up to an absolute magnitude, M$_r$ = -14 mag).}
	\label{tab:example_table2}
	\begin{tabular}{lcccccr} 
		\hline
		  Dwarfs & Giants & Dwarfs & Giants  & Dwarf fraction&\\
		        (full area) & (full area) & (inside R$_{200}$) & (inside R$_{200}$)&   (inside R$_{200}$)             &\\          
		\hline
	       1177 & 168  & 354 & 71 &0.83 \\

		\hline
	\end{tabular}

\begin{tablenotes}
\small
\item\hskip -\fontdimen2{Notes}:  Columns 1 and 2 indicate the numbers throughout the full area covered by the catalogue. Columns 3 and 4 indicate the numbers within R$_{200}$. The last column indicates the dwarf fraction measured within R$_{200}$ of the cluster.
\end{tablenotes}
\end{table*}

\begin{figure*}
    \centering
    \includegraphics[width=1\linewidth]{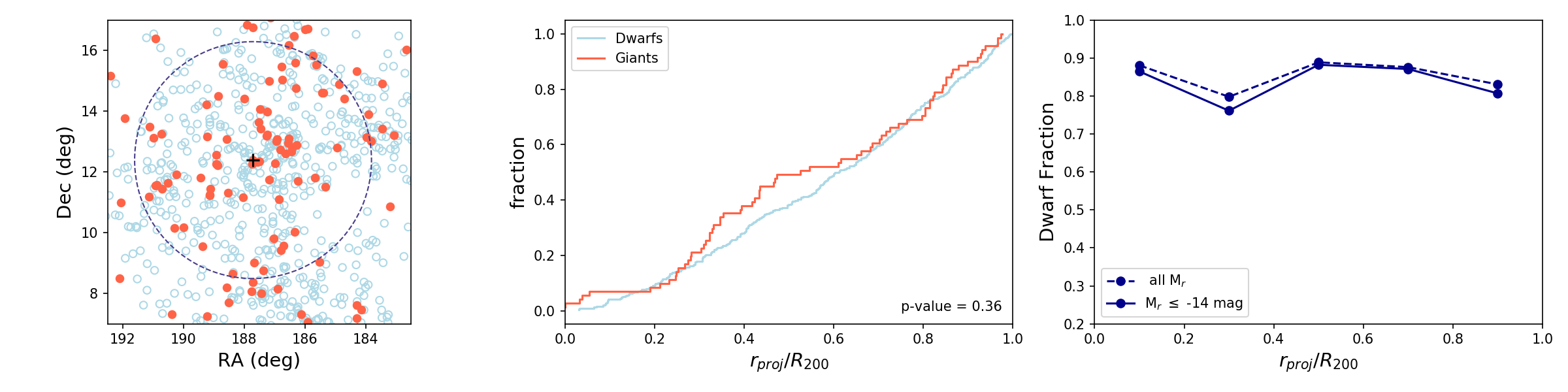}
    \caption{Left panel: Projected spatial distribution of dwarfs (blue open symbols) and giants  (red filled  symbols) from the EVCC catalogue. Black cross indicates the position of M87 galaxy. Dashed circle indicates  the R$_{200}$ region of Virgo cluster. Middle panel: Cumulative distributions of the clustercentric distance of giant and  dwarf  galaxies  inside R$_{200}$ of Virgo cluster.   Note that in both panels the galaxy distribution is limited to M$_r$ = -14 mag. Right panel: Dwarf fraction (DF, measured in bins of distance) as a function of projected cluster-centric distance, measured for the EVCC catalogue without magnitude threshold (dashed line) and for galaxies up to an absolute magnitude, M$_r$ = -14 mag (filled line). }
    \label{fig:fig2}
\end{figure*}

\subsection{Selection of dwarf and giant galaxies}

In the current analysis, we make use of different catalogues since they cover different areas around the Virgo cluster. For the EVCC and NSA-Castignani catalogues our definition is based on the $r$-band, so we have considered faint (“dwarf”) galaxies  those galaxies with -19 $\leq$ M$_r$ $\leq$ -14 mag. The faint limit is imposed to ensure the completeness of the catalogue (as shown in Fig. \ref{fig:fig1}). Bright (“giant”) galaxies were defined as those with M$_r$ $<$ -19.0 mag\footnote{The cut in the absolute magnitude (M$_r$ $=$ -19 mag) is assuming also that all galaxies are at the same distance of Virgo cluster. For example,  the selection for dwarf and giant galaxies using the EVCC catalogue corresponds to a limit in stellar mass log(M$_*$) $\sim$ 9.3 M$\odot$ following the colour-mass-to-light ratio defined in \cite{Roediger2015} assuming that all galaxies are at the same distance as the Virgo cluster (m-M = 31.09 mag, \citealt{Mei2007}).}.  This luminosity-based limit follows the historical convention of classifying dwarf galaxies as those with $M_B > -18$ mag \citep{Binggeli1988}, assuming a typical colour of $B-r \sim 1.0$ mag for such galaxies. Similar to previous studies that rely solely on luminosity to distinguish dwarfs from giants \citep[e.g.,][]{Popesso2005,Rude2020},  hereafter we refer to faint and bright galaxies by the names “dwarfs” and “giants”, respectively.

For the catalogue from \cite{Makarov2011} that comes from the Hyperleda database, our selection of bright and dwarf galaxies was done using the $K$-band magnitude. Therefore, we define dwarfs as those with M$_K$ between -21 mag and -17 mag. Similarly, this faint limit is imposed to ensure completeness of the catalogue (as shown in the third panel in Fig. \ref{fig:fig1}). We define giant galaxies as those with $K$-band magnitude brighter than -22 mag\footnote{We did not counted galaxies  between -21  and -22 mag to avoid possible contamination with our selection of dwarf and giant galaxies. We performed a preliminary test by cross-matching galaxies previously classified as dwarfs and giants based on $r$-band photometry with K-band data. This revealed a minor overlap of galaxies within the specified $K$-band range.}. We highlight that this selection done in the $K$-band is not the same as is done in the other catalogues when we used the $r$-band for the selection; in addition, the completeness (in magnitude) of the catalogues is different, as well as the number of galaxies in each region. Consequently, we expect the results and trends found in this catalogue to be similar to the others, but not exactly the same, and therefore we do not intend to compare them directly.

 Although it is common to use the DGR quantity (number of dwarf galaxies divided by the number of giant galaxies) to compare the spatial distribution of dwarf and bright giant galaxies, when one wants to measure this ratio locally in different regions of the cluster, it makes more sense to estimate the dwarf fraction (DF) as the ratio between the number of dwarf galaxies (Nd) and the number of giant galaxies (Ng) plus dwarf galaxies, DF = Nd/(Ng + Nd). This is to avoid division by zero because there are regions populated mainly by dwarfs without any giant galaxies, as we will show later; henceforth, we mainly use this quantity.

In Table \ref{tab:example_table2} we show  the number of giant and dwarf galaxies, as determined by our criteria for the EVCC catalogue, considering both the entire field of view and the region within the cluster's R$_{200}$ radius\footnote{ The R$_{200}$ radius used  in this analysis corresponds to 3.9 deg. \citep{Arnaud2005,Urban2011}.}. We also include the dwarf fraction (DF). The DF value we get inside the R$_{200}$ radius when using the entire magnitude range of the catalogue becomes 0.85, whereas it is 0.83 when limiting the catalogue to M$_r$ = -14 mag (limit to ensure the completeness of the catalogue). Note that the DF value we are estimating here is a `global' measurement, and although it gives us a measure that helps us to know more about the number of dwarfs with respect to giant galaxies, it does not give us much information about how they are distributed in clusters.

\begin{figure*}
    \centering

    \includegraphics[width=1\linewidth]{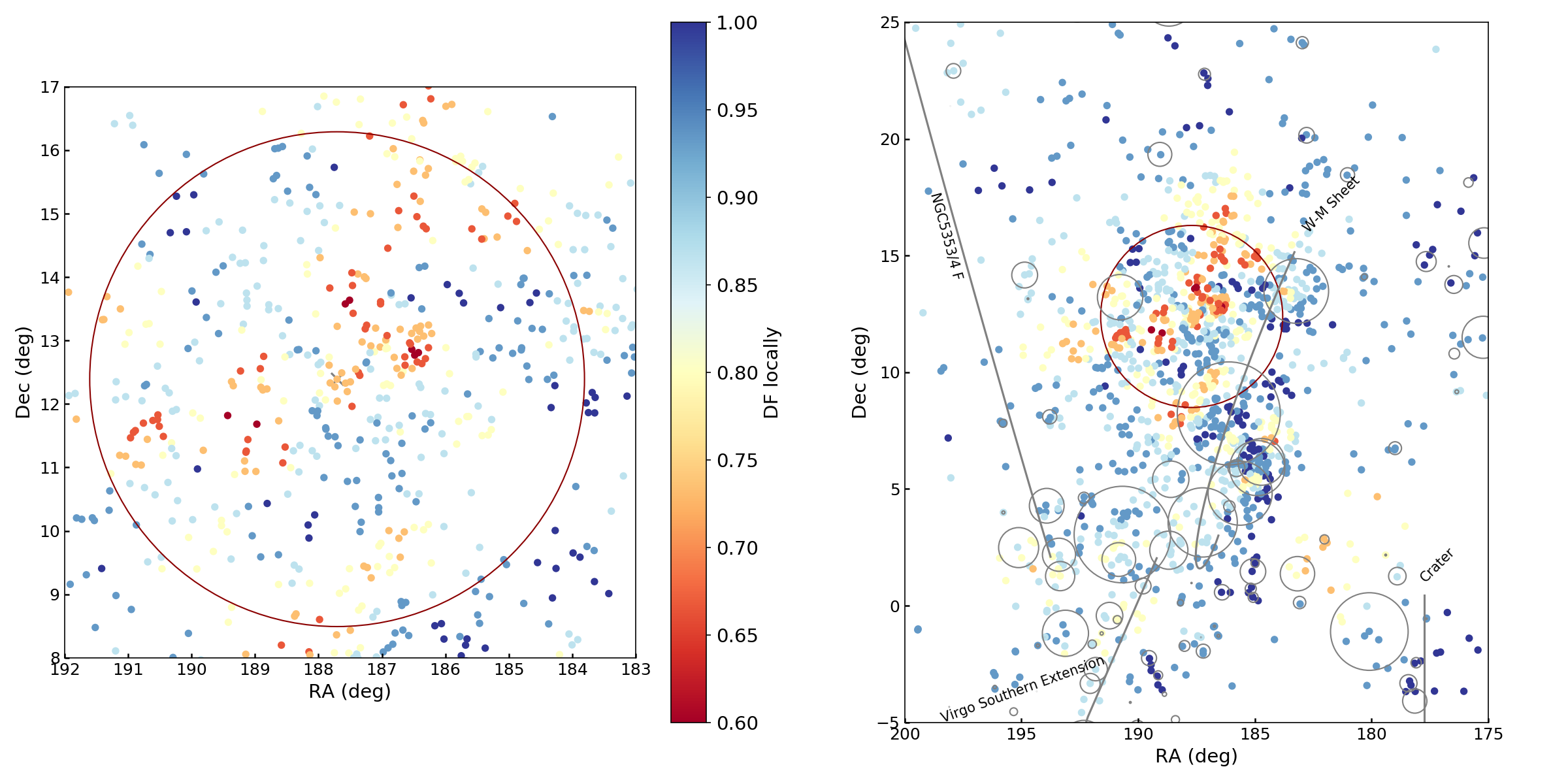} 
    \caption{Galaxy spatial distribution in the EVCC area.  Red circle highlights the R$_{200}$ radius of the Virgo cluster.  Left: DF in the cluster region. Right: DF in the  entire field of view of the EVCC catalogue.  Grey circles highlight  the groups from the galaxy group catalogue of \citet{Kourkchi2017}. The northern grey lines highlights  the filaments of the \citet{Castinagni2022} catalogue. The two filaments in the south were extracted from  Fig. 2 of \citet{Castignani2022a}. The colour bar in  both panels denotes the dwarf fraction (DF) associated with each galaxy, as quantified by considering its fifteen nearest neighbours.}
    \label{fig:fig3}
\end{figure*}

\section{Results}
\label{sec:results}
\subsection{Distribution of giants and dwarfs in the Virgo cluster}
\label{sec:sec3.1}

The analysis of the spatial distribution of galaxies could not only give us information about how they are distributed in and around the clusters but also tell us more about the assembly history of clusters and so help us to better understand the galaxy evolution. Thus, in Fig. \ref{fig:fig2} we show the spatial projected distribution of the giant and dwarf population in the Virgo cluster up to 1 R$_{200}$. Visually inspecting the distribution of dwarfs and bright galaxies in the figure,  we can see that in the left panel, the giant bright galaxies are widely spread across the cluster similarly to the dwarf population, and it is also seen in the cumulative function of the cluster-centric distances shown in the middle panel. In fact,  by taking a Kolmogorov-Smirnov (KS) test between the dwarf and giant distributions, in Virgo both radial distributions are statistically similar (p-value = 0.56, and 0.36 when limiting the catalogue to M$_r$ = -14 mag). Furthermore, when we measure the dwarf fraction in concentric radius bins of 0.2 R$_{200}$ (right panel of the figure) across the cluster, it does not change much, the values remain in 0.7-0.9 (when DF is measured using the entire catalogue and when it is limited to the completeness magnitude limit, M$_r$ = -14 mag).  However, we notice that even when dwarf and giant galaxies are widely distributed across the cluster, there are regions with a higher concentration of dwarfs or bright massive giant galaxies. Therefore, one way to better see the distribution of dwarf and giant galaxies is to measure the DF locally in different regions of the cluster.

Thus, in the left panel of  Fig. \ref{fig:fig3}  we show the spatial distribution of the galaxies in the virialised region of the cluster (using the EVCC catalogue up to an absolute magnitude M$_r$ = -14 mag\footnote{We also checked whether the DF measured  changes when  no magnitude limit is applied to the catalogue. The DF values change a bit, but the variations in different locations of RA and Dec are qualitatively similar as seen in Fig. \ref{fig:fig3}. The same analysis was performed for the other catalogues NSA-Castignani and Makarov/Hyperleda (Figs. \ref{fig:fig4} and \ref{fig:fig5}), and the trend of DF remains similar when using the entire range of magnitudes of the catalogue.}) where we calculated the DF in different local regions.  That is, for each galaxy we select its 14 nearest neighbours, thereby defining a “local” region containing 15 galaxies in total. Neighbour selection is based on projected distances measured in two dimensions using the RA and Dec coordinates. The dwarf fraction (DF) is then computed by counting the number of dwarfs (Nd) and giants (Ng) within this local neighborhood and taking the ratio of dwarfs to the total number of galaxies (Nd+ Ng). In this way, every galaxy has an associated projected DF value (shown as colours in Fig. \ref{fig:fig3}).

 We notice that there are several zones with a high concentration of massive giant galaxies (redder-coloured points), with a DF $\sim$ 0.6 meaning that at least $\sim$ 40 \% of the galaxies in that neighbourhood are massive/giant galaxies.  These regions are in fact related to the galaxy substructures found already in this cluster, e.g., M86, M84, M60, and M49. In other regions of the cluster, we observe a predominance of dwarf galaxies with a DF $\geq$ 0.8 (bluer-coloured points).

\subsection{Fraction of dwarf galaxies around Virgo}
\label{sec:sec3.2}

To see how the dwarf fraction changes not only across the cluster but also at greater distances, in the right panel of Fig. \ref{fig:fig3} we show the entire EVCC area that covers a region of $ \sim$ 25x25 deg$^2$. Looking in the surrounding area of the cluster, we notice some patches with a higher concentration of giants that seem to be correlated with the groups falling into the cluster. We also observe that there are regions with a high concentration of dwarfs (DF $\geq$ 0.8), for example, in the lower right region outside the cluster and,  in some cases, not even giants. One possible explanation for this is that they are galaxies in groups composed mainly of dwarf galaxies. A better understanding of how the DF changes at a larger scale could help us understand how the DF changes inside the Virgo cluster; for instance, we could have an idea on how the surrounding area is feeding the cluster with dwarf and giant galaxies and thus increasing or decreasing the DF inside the cluster. 

\begin{figure*}
    \centering
    
    \includegraphics[width=17cm, height=11.5cm]{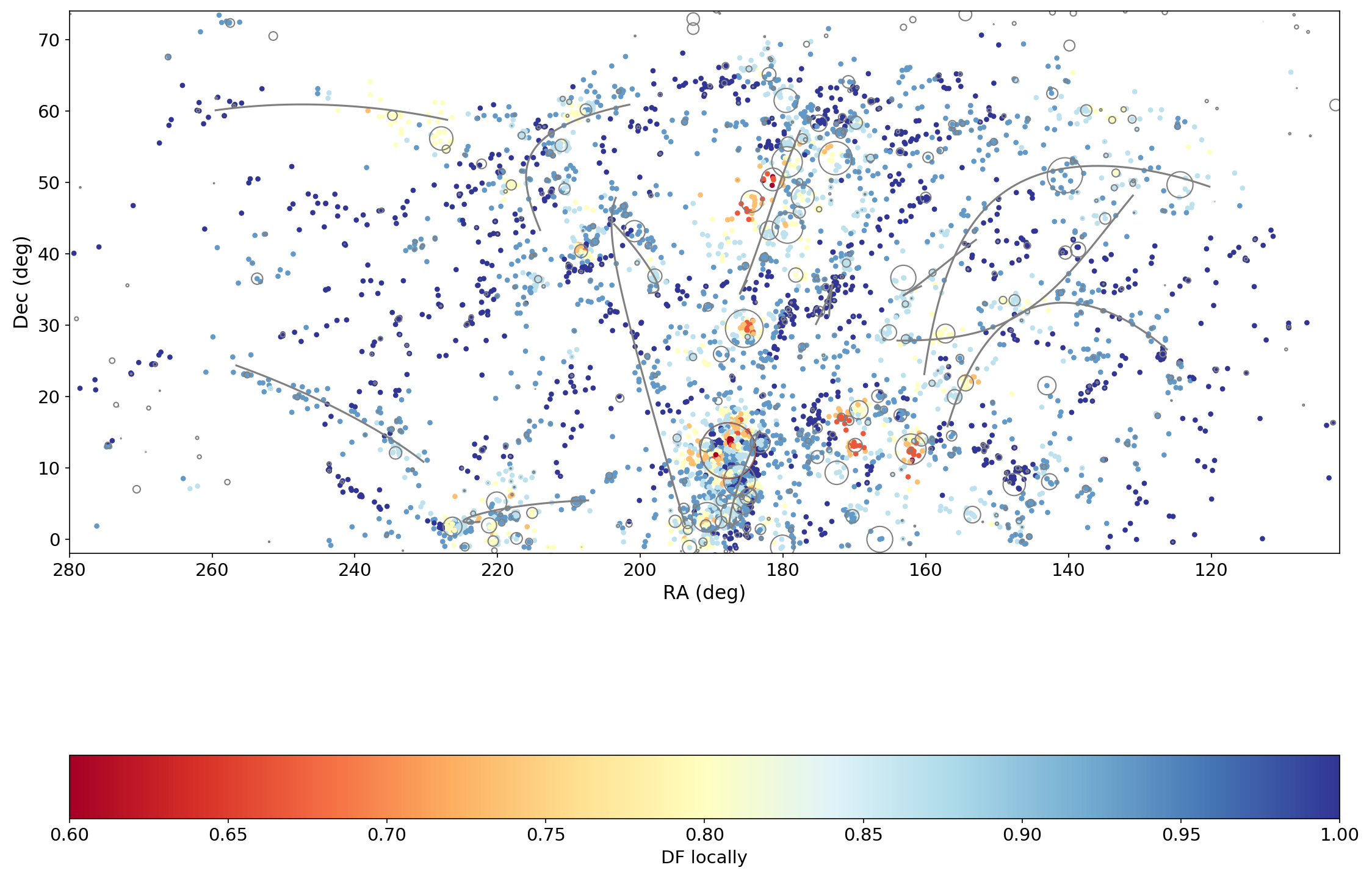}
    \caption{Galaxy spatial projected distribution  in the large scale structure around Virgo (northern hemisphere) from the NSA-Castignani catalogue \citep{Castinagni2022}. Colour indicates the local DF  measured on each galaxy in the region enclosing its fifteen near neighbour galaxies.  Red circle highlights the R$_{200}$ radius of the Virgo cluster, while grey circles highlight the (loose and rich) groups from the galaxy group catalogue of \citet{Kourkchi2017} (in the  velocity range 500 $<$ $v_r$ $<$ 3300 km/s; the circle sizes correspond to the  projected virial radius of the groups described in that work). The grey lines highlights  the filaments of the \citet{Castignani2022a,Castinagni2022} catalogue; Leo Minor F, Canes Venatici F, Bootes F, Ursa Major Cloud, Leo II B F, Leo II A F, Virgo III F , Leo Minor B F, W–M Sheet, NGC 5353/4 F, Serpens F, Draco F, and Coma Berenices F.}
    \label{fig:fig4}
\end{figure*}

In Fig. \ref{fig:fig4} we show the DF calculated using the NSA-Castignani galaxy catalogue. This catalogue covers a large area around Virgo (but only in the northern hemisphere, $\sim$ 60 degrees in declination and around 100 degrees in right ascension, as shown in the figure), where we explore the DF across it, for example, in the various filaments around Virgo reported in previous works \citep[e.g.,][]{Castinagni2022}. Similarly as we did for the EVCC we limited the sample to DF estimation up to M$_r$ = -14  mag (to ensure completeness of the catalogue).  As can be seen in the figure, in some filaments there are regions with a relatively high concentration of giant galaxies (as might be expected in dense regions); see for example the filament in the northern zone; known as the Ursa Major cloud as reported in the literature. However, also note that there are other filaments with a fairly large fraction of dwarf galaxies; see, for example, the `small' filament in (RA, Dec) $\sim$ (170, 35) degrees, which appears to be composed mainly of dwarf galaxies (DF $>$ 0.9). This could be because this filament is relatively new and/or the dwarf galaxies in these filaments belong only to small galaxy groups. In fact, a number of studies have reported groups with only dwarf galaxies \citep[]{Stierwalt2017}. 

By visually examining this figure, we can gain a qualitative understanding of the dwarf fraction's variation across different environments. For example, there is a higher concentration of dwarfs in filaments compared to the cluster, and a high fraction of dwarf galaxies in distant regions from the cluster and filaments. Note also that the groups in filaments also have different fractions of dwarf galaxies. A more detailed comparison between the different environments will be further explored in the next section.  In addition, as a consistency check,  in the Appendix \ref{section_app1}  we present measurements of the local DF in Cartesian supergalactic coordinates, analogous to those shown in Fig. \ref{fig:fig4}, while also accounting for the influence of local attractors \citep{Mould2000, Castinagni2022}. We find no significant differences compared to the local DF obtained in RA–Dec coordinates.

\begin{figure*}
    \centering
  
    \includegraphics[width=17cm]{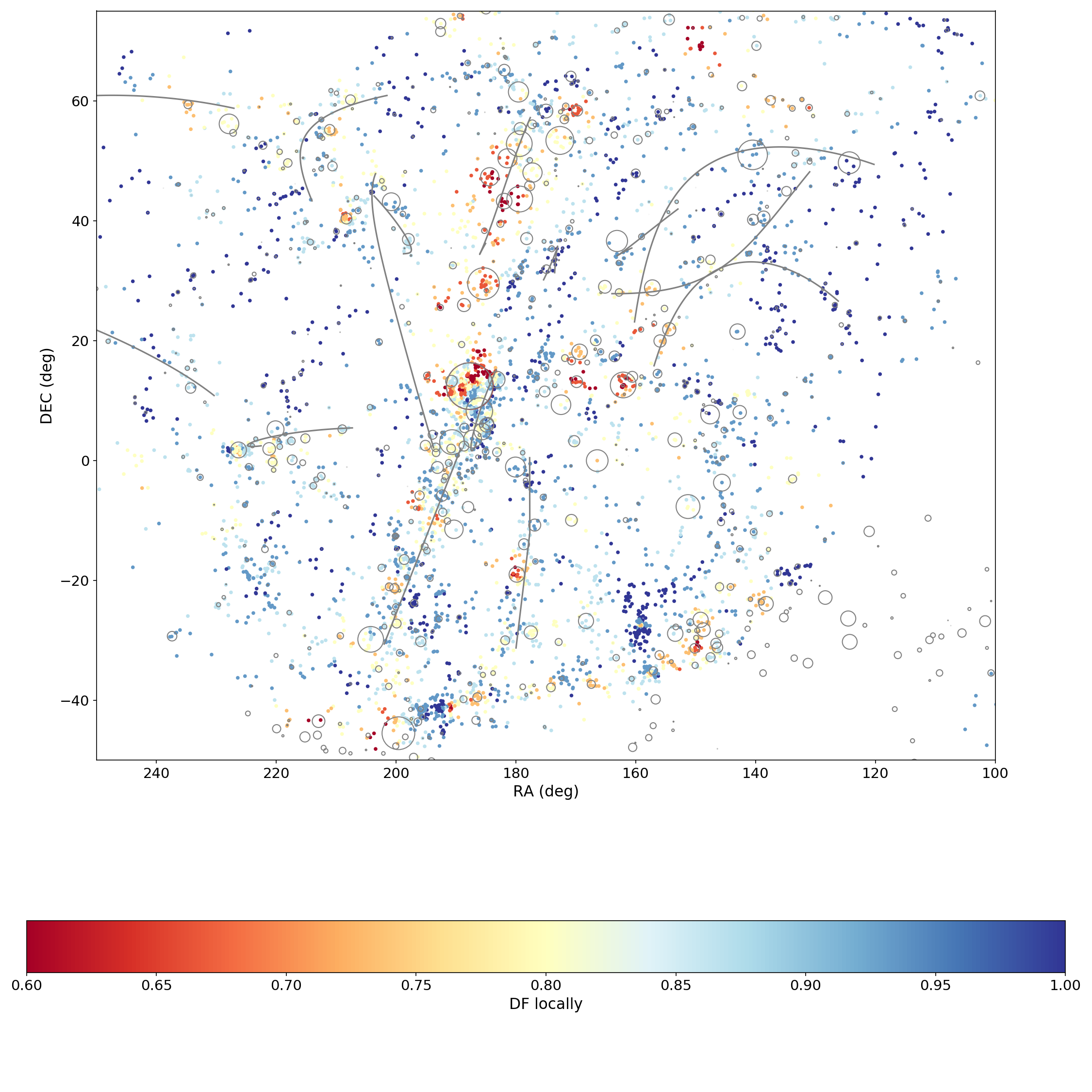}
    \caption{Galaxy spatial projected distribution  in the large scale structure around Virgo (including the southern hemisphere) from the Hyperleda catalogue described in \citet{Makarov2011}. Colour indicates the DF on each galaxy  measured in the region  containing the fifteen near neighbour galaxies.  Red circle highlights the Virgo cluster,  while grey circles highlight the (loose and rich) groups from the galaxy group catalogue of \citet{Kourkchi2017} (in the  velocity range 500 $<$ $v_r$ $<$ 3300 km/s). The grey lines highlights  the filaments of the \citet{Castinagni2022} catalogue. Note that all the filaments reported in \cite{Castinagni2022} are plotted the figure; Leo Minor F, Canes Venatici F, Bootes F, Ursa Major Cloud, Leo II B F, Leo II A F, Virgo III F , Leo Minor B F, W–M Sheet, NGC 5353/4 F, Serpens F, Draco F, and Coma Berenices F. The two filaments in the south: Virgo Southern Extension and Crater were extracted from \citet[][see their Fig. 2]{Castignani2022a}.}.
    \label{fig:fig5}
\end{figure*}

Although with the NSA-Castignani catalogue we can explore a huge area around Virgo, we also aimed to check how the DF locally changes in the southern region. To do this, we use the catalogue provided in \cite{Makarov2011} (which is based on the Hyperleda database). However, given that this catalogue provides photometry in the $K$-band only, we made the selection of the dwarf and giant bright galaxies using this filter.  Therefore, we might expect a similar trend in the results as observed with the other catalogues, but not necessarily the same since the DF values measured now are based on this $K$-band catalogue.

Thus, in Fig. \ref{fig:fig5} we now show the distribution of DF in both hemispheres around Virgo\footnote{One caveat to note is that the local number of galaxies (Ng + Nd)  using this $K$-band catalogue may not represent the total population, since objects within the range -22 $<$ K $<$ -21 mag  have been discarded, as explained in Sect. \ref{sec:data}. However, we emphasise that our goal here is not to quantitatively compare these findings with results from other catalogues where $r$-band selection was used.}. The \citet {Makarov2011}  catalogue covers a much larger area;  however, here we show only the area surrounding Virgo, covering the area where all filaments previously found that could be associated with the Virgo cluster (see the lines in the figure and also \citealt{Castinagni2022,Castignani2022a}). Note, for example, that in the dense (if not the densest) filament in the southern region, low values of the DF are measured, meaning a significant fraction of giant galaxies inhabit that area. In fact, we see regions with a DF $\sim$ 0.5 - 0.7 meaning that at least the 30\% or more of the galaxies in that filament are brighter massive ones which seem to be related with the several groups on that filament (see grey circles in the figure). Similarly, we also see in the most southern region (border) a relatively high fraction of bright giant galaxies which seem to be correlated with the several groups reported in that area. On the other hand, we again note places populated by higher concentrations of dwarf galaxies see, for example, the region around (RA, Dec) $\sim$ (160,-30), which also in some cases seem to be related with the groups identified. Additionally,  we note that, at least by eye,  low-density regions not associated with groups or filaments tend to be bluer, meaning higher concentration of dwarfs. 

Therefore, looking at the large-scale structure in this figure and in Fig. \ref{fig:fig4}, we see that the DF changes in different environments, the dwarf fraction does not have the same value; it is different in filaments, in the cluster region, and even in the groups located all over the place (grey circles in the figure). We will discuss further what might produce these changes in the DF later in Sect. \ref{sec:summary}.

\subsection{DF in different environments}
\label{sec:sec3.3}

 To check how the dwarf fraction changes in different environments, we measure the distribution of the DF we estimated locally as shown in Figure \ref{fig:fig4}. Three distinct environmental regions were defined: a high-density region, represented by the Virgo cluster (up to 1R$_{200}$); an intermediate-density region, comprising the surrounding filamentary structures; and low-density regions, corresponding to areas not included in the previously defined environments. We use the filament catalogue of \citet{Castinagni2022}, which contains thirteen filaments in the northern hemisphere surrounding the Virgo cluster, as highlighted in Fig. \ref{fig:fig4}.

 Thus,  in Fig. \ref{fig:fig6} (left panel) using the galaxy NSA-Castignani catalogue we show the distributions of the DF inside the cluster region (orange box), in the filamentary structure (green box), and in the remaining low-density regions\footnote{The galaxies belonging to this region are selected as those not belonging to the filamentary structure  neither the ones detected in the cluster area.} (blue box).  Overall, when comparing the three environments, we see that they span a wide range of DF values, as shown in Fig. \ref{fig:fig4}, indicating that each environment contains local zones contributing both low and high DF values. In the Virgo cluster region, however, the DF range extends slightly toward lower values compared to the filaments (see the peak in the distributions, the median values being 0.87 and 0.93, respectively). Since the Virgo cluster is a very dense region with several substructures falling into it, we expect an increase in the number of giant galaxies and, consequently, a decrease in the measured dwarf fraction relative to the filaments.  On the other hand, Fig. \ref{fig:fig4} qualitatively shows that some dense filaments also exhibit a higher contribution from massive galaxies (similar DF differences are also observed from group to group). In regions we classified as “low density”, although they span a DF range comparable to the other two environments, the contribution from local zones with higher DF values is slightly greater. This results in a marked concentration of dwarf galaxies, with DF values predominantly in the range 0.8–1.0 (median value being 0.93), as shown in Fig. \ref{fig:fig6}. To complement this aspect, we also checked the spatial distribution of galaxies in Fig. \ref{fig:fig4} that have a very high DF ($\geq$ 0.95) and a relatively low DF ($\leq$ 0.75) and observed a similar trend on the distribution of galaxies in different environments (see Fig. \ref{fig:fig3_appen} in the Appendix \ref{section_app2}). The galaxies with high DF are widely distributed in all environments with a predominance in less dense environments, whereas galaxies with low DF are concentrated in dense regions such as the cluster and in some filaments.

 In addition to the analysis in these three environments for completeness, we conduct a further analysis of the dwarf fraction (DF) variation within the filamentary environment, specifically through a comparative analysis of results derived from the exclusion of galaxy groups with those derived from the exclusive consideration of galaxy groups in the filaments (bottom panel of Fig. \ref{fig:fig6}). Galaxies belonging to groups were selected using the catalogue from \citet{Kourkchi2017}. As can be seen, the groups in filaments show a lower DF as they are contributing with more massive galaxies than filaments without galaxy in groups (see the lime box becoming more wide and extended to the left side, where the DF values are lower). Furthermore, we also checked the comparison in different environments using the Hyperleda catalogue \citep{Makarov2011} (in the northern area) and found similar trends (see the right panel in Fig. \ref{fig:fig6}). Similarly,  we also find similar trends when using the filamentary structure we detect with 1-DREAM \citep{Canducci2022} as shown in the Appendix \ref{section_app3}.

As a final step to complement our analysis of the dwarf fraction (DF) in various environments, we briefly checked DF variations within galaxy groups reported by \citealt{Kourkchi2017} (groups covering the area shown in Fig. \ref{fig:fig4}). We calculated the mean local DF for each group by averaging the local DF values of its member galaxies. The resulting mean local DF values increase from approximately 0.6 to 1, consistent with the finding shown in Fig. \ref{fig:fig6} for the groups.

\begin{figure*}
    \centering
    \includegraphics[width=1\linewidth]{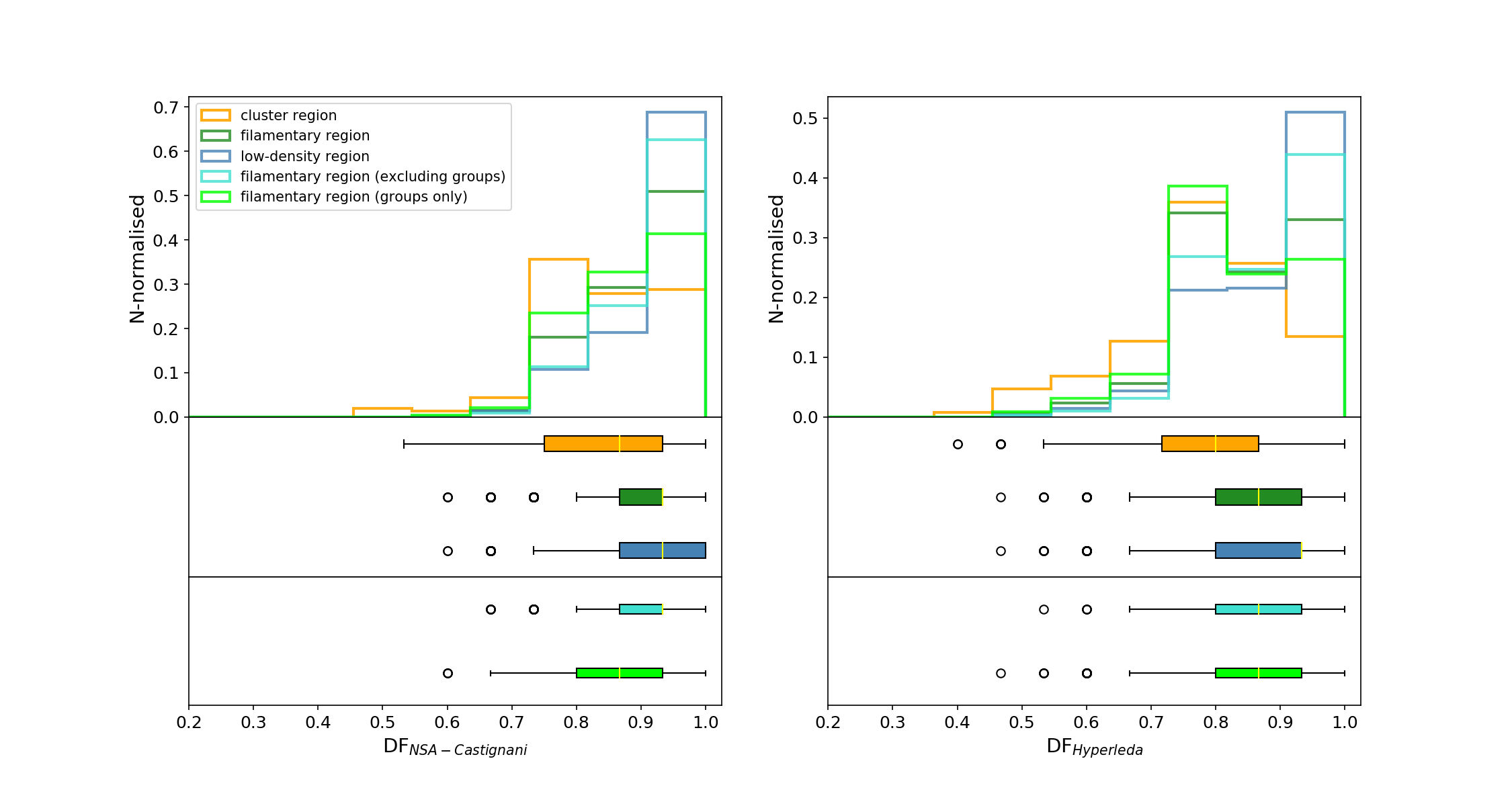}
    
    \caption{DF distribution in different environments using the NSA-Castignani catalogue (left) and Hyperleda catalogue (north hemisphere; right). Top panel: Distribution of DF values in the Virgo cluster region (orange), filamentary region (green), and low-density regions (blue). Additionally, we include the DF distribution of the filamentary region excluding galaxies belonging to groups (cyan) and vice versa, including only galaxies belonging to groups (lime). Bottom panels: Box plots of the same distributions.  Note that the box covers the first and third quartile of the data (e.g., between the 25 and 75 percentiles), while the line above it indicates the median value. The whiskers indicates the lowest and highest value still within 1.5 times the interquartile range, and the open dot symbols mark the outliers.}
    \label{fig:fig6}
\end{figure*}

\section{Discussion and Conclusions}
\label{sec:summary}

The present study is centred on the Virgo cluster, a nearby cluster situated within a highly enriched and dense environment, thereby constituting an advantageous location for examining the spatial distribution of dwarf and giant galaxy populations within the cluster and its surrounding regions. The main findings of this study are summarised as follows.

\begin{itemize}
\item Even when in general the number of dwarfs is larger than the giant galaxies inside the Virgo cluster (inside of 1 R$_{200}$ the DF $\sim$ 0.8), this does not mean that all regions in the cluster will have this value. In fact, there are local variations of the DF (Figs. \ref{fig:fig2} and left panel of Fig. \ref{fig:fig3}).

 \item The fact that the Virgo cluster is embedded in a rich environment, surrounded by several filaments that constantly supply the cluster with new substructures, leads to local variations in the DF within the cluster (Fig. \ref{fig:fig3}).

   \item  Even inside the cluster and in its surrounding area there are regions with few or no giant galaxies at all. For example, with locally DF between 0.8 - 1.0. Similar giant free regions are also observed at greater distances from the cluster in some filamentary regions and some in low-density regions (Figs. \ref{fig:fig3}, \ref{fig:fig4}, and \ref{fig:fig5}).

    \item When comparing the dwarf fraction in different global environments, we find that the contribution of local regions with higher DF values is greater in low-density regions. This contribution tends to decrease in more dense environments, such as the filaments and within the Virgo cluster. A comparison between the filaments and the cluster area shows that local zones with high DF values are slightly more present in filaments. However, there are also variations in the DF from one filament to another, depending on the presence of galaxy groups (Figs. \ref{fig:fig6} and \ref{fig:fig3_appen}).

\end{itemize}

In Sect. \ref{sec:sec3.1} we saw that when we measure the total fraction of dwarf galaxies inside the R$_{200}$ radius of the Virgo cluster it has a value of $\sim$ 0.8, meaning a high number of dwarf galaxies compared to massive galaxies, and this fraction does not change much when going farther from the cluster centre, similarly to found in previous studies \citep[e.g.,][]{Sabatini2003}. However, even though the trends are similar, a quantitative comparison with other studies is not simple because of the different definitions of dwarf and giant galaxies used (-19 $\leq$  M$_r$ $\leq$ -14 mag and M$_r$ $<$ -19  mag for the dwarf and giant definitions used in this work). Similarly, because most previous studies have focused on measuring the dwarf fraction within a cluster's virial radius or at relatively small distances from it, our results, which estimate the DF in local zones much farther from the cluster, such as in filaments and field regions, cannot be compared fairly.

On the other hand, even when dwarf and giant galaxies are widely distributed inside the cluster,  we saw that the DF actually changes in different regions of the cluster and in its surrounding area. There are regions with higher concentrations of giant galaxies, which in fact correlate with the groups falling into the cluster and thus contributing massive galaxies. In fact, looking at the surrounding area, there are several filaments feeding the cluster with groups that eventually will contribute giants and dwarfs galaxies and so produce changes in the DF in different regions of the cluster.  At the same time, we might consider that, even though the number of dwarfs is in general larger in the cluster, they are more exposed to disruption by tidal forces and so one might expect some lower DF values in inner regions of the clusters. Therefore, in some ways, there is `competition'  between the gravitational tidal effect decreasing the DF and the feeding of galaxies coming from the filaments that might tend to increase the DF.

However, on the other hand,  even in dense regions such as in the cluster or in the near surrounding area there are some regions with a high concentration of dwarf galaxies only which seem to be not correlated with any giant galaxy. Certainly, we have to keep in mind that we are looking at a projected distribution, but this does not mean that it is not important to understand those regions. In fact, these dwarf galaxies that seem to be not associated to giants in the Virgo cluster were already reported \citep{Ferguson1992,Sabatini2003}.  One possible explanation could be that they are groups composed mainly of dwarf galaxies, as has already been reported in the literature \citep[e.g.,][]{Makarov2012,Stierwalt2017}. On the other hand, they could be the remaining groups members where the more massive galaxies sank to the densest region due to dynamical friction or they are dwarf galaxies that have been liberated from their former groups when their group was disrupted \citep[][]{ChoqueChallapa2019,Benavides2020,Haggar2023}.  This fact is even more interesting to analyse when we look also at further distances from the cluster, for example, in the filamentary structure surrounding it. One useful approach that could help to understand the physical processes shaping these regions with only dwarf galaxies would be the use of hydrodynamical cosmological simulations; for instance, we could track those regions in time and see how they became regions with few or no giant galaxies at all, or vice versa, regions with higher concentration of giants. We will further analyse these aspects in a follow-up study.

Also, another interesting aspect we can see when we look at the DF measured on a large scale (Sections \ref{sec:sec3.2} and \ref{sec:sec3.3}) is the fact that even when the dwarf fraction is higher in the filamentary region compared to the Virgo cluster, the DF changes from filaments to filaments (and so from group to group). Additionally, another interesting point to note is related to how dwarf and giant galaxies might fall into the cluster, because it seems to be that there are several regions with a high concentration of dwarfs that are not associated with groups or filaments. Overall, our results remain qualitatively consistent when adopting supergalactic coordinates (Appendix \ref{section_app1}). In the present analysis, we mainly studied the changes in the local dwarf fraction in the Virgo cluster and the large scale around it. However, in future work, we aim to compare our findings with those from clusters exhibiting different properties (e.g., dynamical state, surrounding environment).


\begin{acknowledgements}

We thank the referee for their useful comments and suggestions, which have contributed to improving the paper. NCC acknowledges a support grant from the Joint Committee ESO-Government of Chile (ORP 028/2020) and the support from FONDECYT Postdoctorado 2024, project number 3240528. RS acknowledges financial support from FONDECYT Regular 2023 project No. 1230441 and also gratefully acknowledges financial support from ANID - MILENIO
NCN2024\_112. JALA acknowledges support from the Agencia Estatal de Investigaci\'on del Ministerio de Ciencia, Innovaci\'on y Universidades (MCIU/AEI) under Grant `WEAVE: EXPLORING THE COSMIC ORIGINAL SYMPHONY, FROM STARS TO GALAXY CLUSTERS' and the European Regional Development Fund (ERDF) with reference PID2023-153342NB-I00/10.13039 501100011033.
    
\end{acknowledgements}

  \bibliographystyle{aa} 
   \bibliography{referencias} 

\begin{appendix}


\clearpage

\clearpage
\FloatBarrier

\begin{figure*}[t!]
\centering
\begin{minipage}{0.95\textwidth}
\begin{multicols}{2}

\section{DF locally measured with Cartesian Supergalactic Coordinates}
\label{section_app1}

In Figure \ref{fig:fig4}, we measured the local DF in regions enclosing fifteen near neighbours. The distances to identify these neighbors were calculated in 2D using projected RA and Dec coordinates. Additionally, here as a check, we perform the same analysis but in supergalactic coordinates. In Figure \ref{fig:fig1_appen}, we measure the local DF similarly to the equatorial coordinates, but the distances were calculated using 3D Cartesian Supergalactic Coordinates (SGX, SGY, SGZ). We also used a cosmic velocity model measured in \citet{Castinagni2022} for the calculation of these coordinates to account for the effect of attractors in the local universe \citep{Mould2000}. Qualitatively, this figure and Fig. \ref{fig:fig4} are similar. We also did a quantitative comparison  between them, this is,  comparing the DF   value galaxy to galaxy. We find  that  the difference is small for most of the galaxies ($\sim$80 \%) in the sample having a difference  $\leq$ 0.08  (within 1$\sigma$) of the distribution   as  can be  seen in Fig. \ref{fig:fig2_appen}.

\columnbreak


\begin{center}
\addtocounter{figure}{1} 
\includegraphics[width=\linewidth]{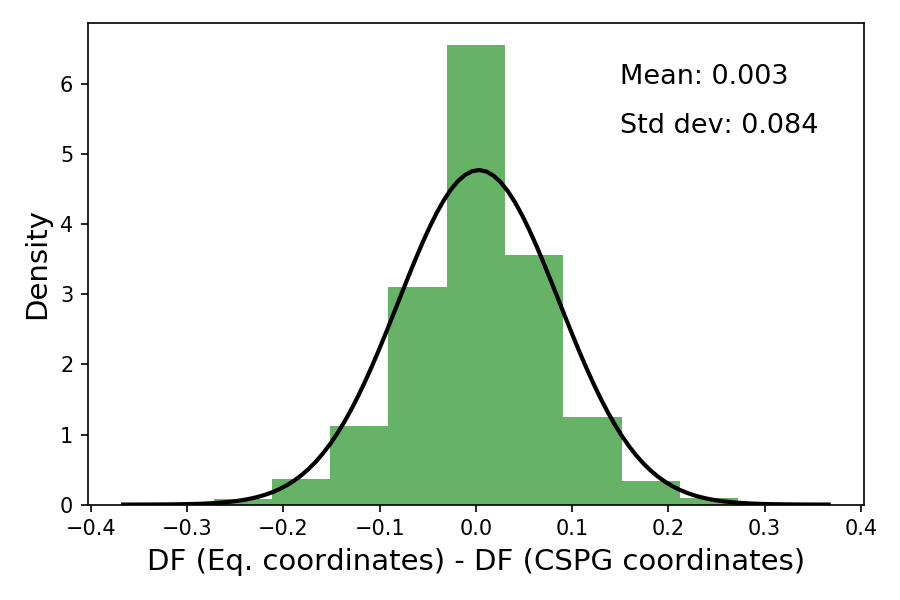}
\caption{Difference between the DF measured on each galaxy with 2D coordinate system and  with the Cartesian supergalactic coordinates. Black line highlights a Gaussian distribution.}
\label{fig:fig2_appen}
\end{center}

\end{multicols}

\vspace{1.2em}
\begin{center}
\addtocounter{figure}{-2} 
\includegraphics[width=\textwidth]{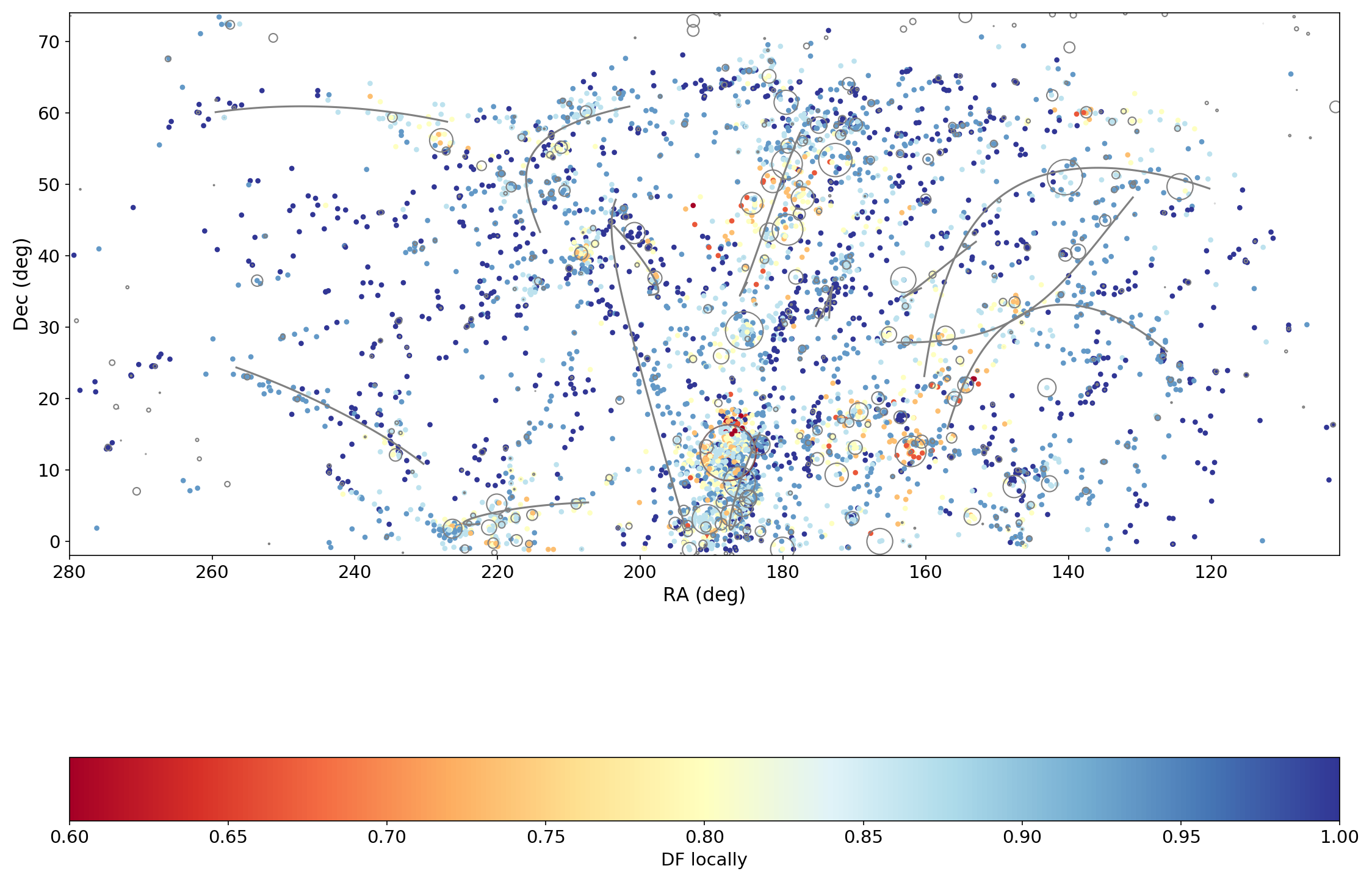}
\caption{Similar to Fig. \ref{fig:fig4}. Projected spatial distribution of galaxies in the large-scale structure around Virgo (NSA-Castignani catalogue), but calculated  using Cartesian Supergalactic Coordinates (SGX, SGY, SGZ).}
\label{fig:fig1_appen}
\end{center}

\end{minipage}
\end{figure*}

\FloatBarrier
\clearpage

\newpage
\clearpage
\section{Distribution of galaxies with  high and low DF}
\label{section_app2}

\begin{figure}[ht!]
    \centering
    \hspace*{1cm}\includegraphics[width=0.85\textwidth]{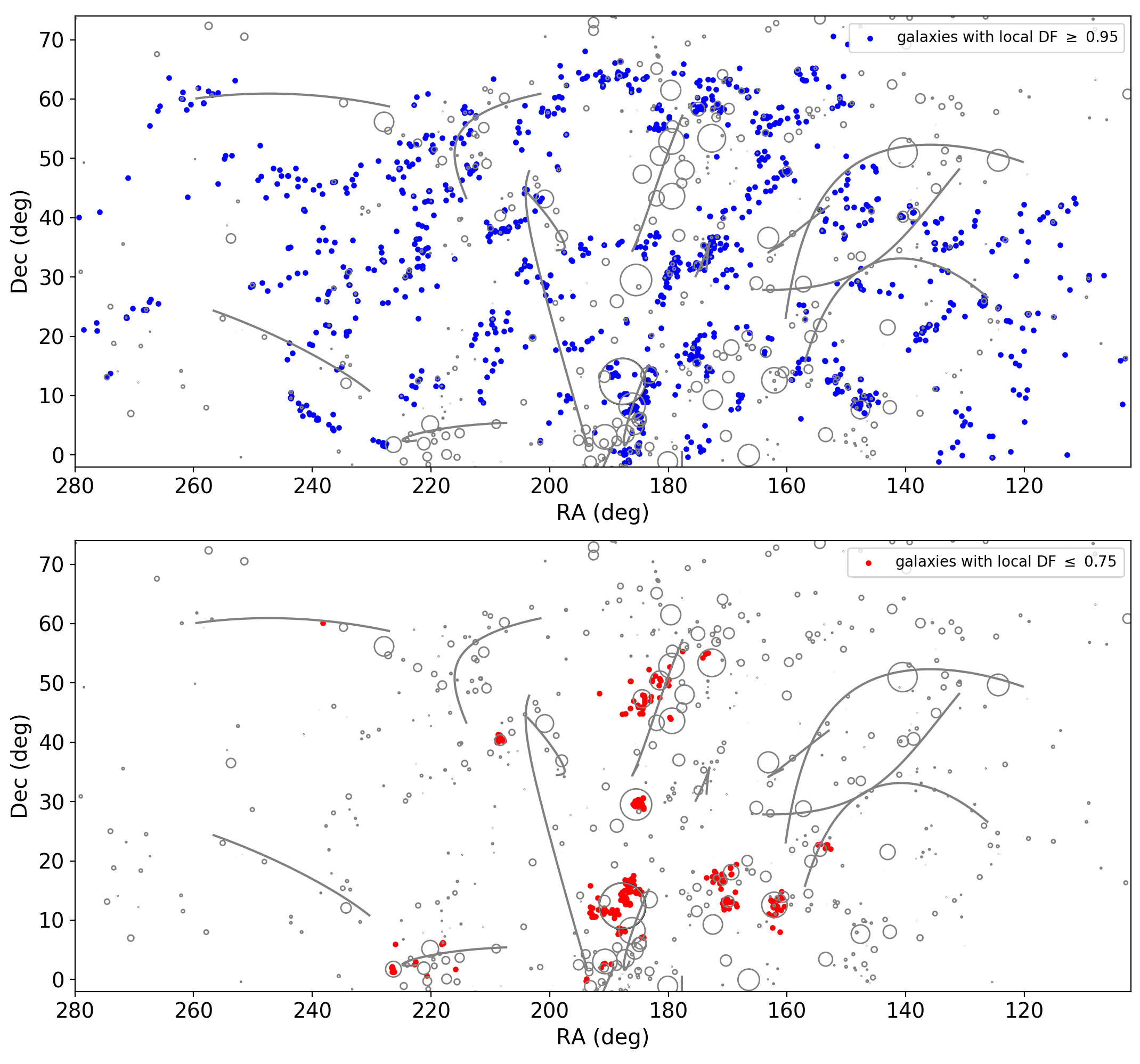}
    \onecolumn
    \caption{ Distribution of galaxies with a DF $\geq$ 0.95 (top panel) and with a DF $\leq$ 0.75 (bottom panel).}
    \label{fig:fig3_appen}
\end{figure}

In the top panel of Fig. \ref{fig:fig3_appen} we show the projected spatial distribution  of galaxies with a very high DF ($\geq$ 0.95, regions mainly populated by dwarfs) selected from Fig. \ref{fig:fig4}. Similarly, in the bottom panel we show the distribution of those galaxies with a low DF ($\leq$ 0.75,  region populated by a relatively  high concentration of giants - at least 25\% of them are giant galaxies in that neighbourhood-). From this figure we can see that dwarf galaxies are widely distributed in all environments with a predominance in less dense environments as compared to the bright giant galaxies.

\FloatBarrier

\twocolumn

\section{DF in different environments:
comparison with filamentary structure detected with 1-DREAM}
\label{section_app3}

\begin{figure}
    \centering
    \includegraphics[width=8cm]{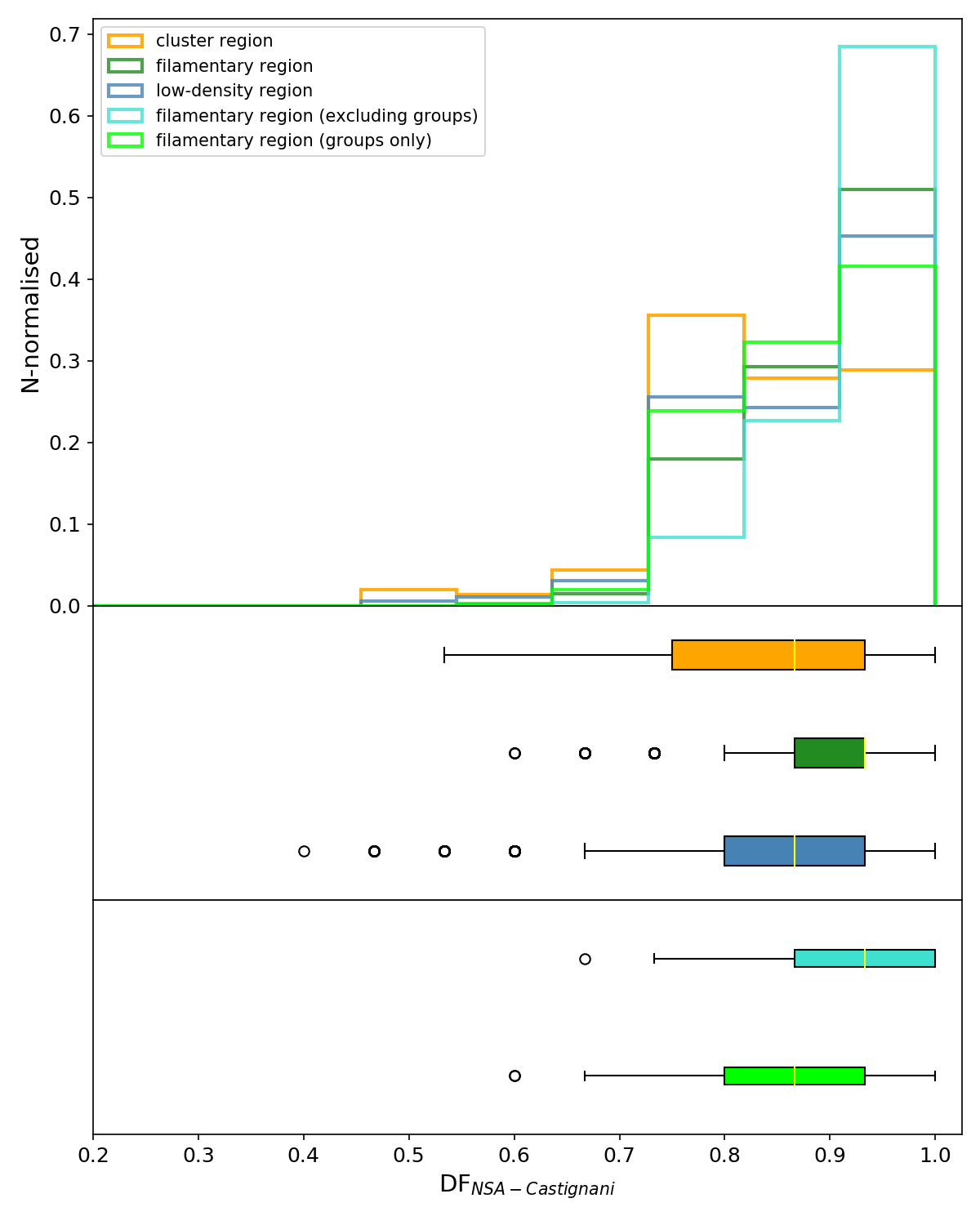}
    \caption{DF distribution in different environments using the NSA-Castignani galaxy catalogue using the filamentary structure detected by 1-DREAM. Top panel: Distribution of DF values in the Virgo cluster region (orange), filamentary region (green), and low-density regions (blue). Additionally, we include the DF distribution of the filamentary region excluding galaxies belonging to groups (from \citet{Makarov2011} group catalogue; cyan) and vice versa, including
only galaxies belonging to groups (lime).}
    \label{fig:fig4_appen}
\end{figure}

We also check the dwarf fraction (DF) distribution found in Fig. \ref{fig:fig6} for the filamentary structure we detect with LAAT/1-DREAM \citep[1-Dimensional Recovery, Extraction, and Analysis of Manifolds ][]{Canducci2022}. 1-DREAM is a toolbox composed of five algorithms ranging from the detection of filaments to the finding of their central spines \citep[see also][]{Awad2024,Raj2024}. For the purpose of this analysis, we only make use of one of its algorithms; the Locally Aligned Ant Technique (LAAT). LAAT uses the principle of the ant colony methodology and so aims to identify and highlights 1-D structures via the local value of the pheromone count. Using a threshold in this parameter, we can filter out noise surrounding the filaments. We run LAAT/1-DREAM on catalogues from NSA and  the one from \citet{Makarov2011} and did several tests, we tried varying the two main parameters of the algorithm, the neighbourhood radius and the pheromone cut. For example, a lower pheromone cut gives less noise removal and larger radius links up structures with gaps that are more extended. We noticed that DF results were not strongly affected by the exact choice of parameters. We used the right ascension, declination, and redshift information from the catalogues as an input parameters, and selected then a minimum pheromone cut of 0.1 from the output data. Thus in Fig. \ref{fig:fig4_appen} we show the DF distribution for NSA-Castignani catalogue using the filaments  extracted by 1-DREAM (green box). As can be noted, the DF distribution is similar using 1-DREAM and the filament catalogue from Castignani (e.g., the median DF value $\sim$ 0.9 for the filamentary region). Additionally, in the bottom panel of the figure we did the same analysis where we exclude and include only the group galaxies, but this time using the  group catalogue of \citet{Makarov2011}. As can be noticed the result remains similar as when we use \citet{Kourkchi2017} group catalogue.

\end{appendix}

\end{document}